# Screening of half-Heuslers with temperature-induced band convergence and enhanced thermoelectric properties

Jinyang Xi,[1,†] Zirui Dong,[2,†] Menghan Gao,[1] Jun Luo,[2] and Jiong Yang[1,*]

[1]Materials Genome Institute, Shanghai University, Shanghai 200444, China.

[2]School of Materials Science and Engineering, Shanghai University, Shanghai 200444, China.

[†]The authors contributed equally.

**Corresponding Author:**

[*]Jiong Yang, Email: jiongy@t.shu.edu.cn

**Abstract**

Enhancing band convergence is an effective way to optimize the thermoelectric (TE) properties of materials. However, the temperature-induced band renormalization is commonly ignored. By employing the recently-developed electron-phonon renormalization (EPR) method, the nature of band renormalization in half-Heusler (HH) compounds TiCoSb and NbFeSb is revealed, and the key factors for temperature-induced conduction band convergence in HH are found out. Using these as the screening criteria, 3 out of 274 HHs (TiRhBi, TiPtSn, NbPtTl) are then stood out from our *MatHub-3d* database. Taking TiPtSn as the example, it shows the conduction band convergence at mid-high temperature, and further resulting in enhanced Seebeck coefficient $S$: *e.g.*, at 600 K with electron concentration $10^{20}$ cm$^{-3}$, the predicted $S$ with and without renormalized band is 352.83 μV/K and 289.52 μV/K, respectively. Herein, the former is closer to our measurement value of 338.79 μV/K. Besides, the effective masses obtained from calculation and experiment are both enlarged with temperature, indicating the existence of band convergence. Our work demonstrates for the first time the significance of adding the temperature effect on electronic structure in the design of potential high-performance TE materials.

**Introduction**

Thermoelectric (TE) materials have attracted enormous attention by the advantages in refrigeration and waste heat recovery.[1,2] The TE performance of a material is characterized by the dimensionless figure of merit $ZT = S^2\sigma T/(\kappa_e + \kappa_L)$, where $S$ is the Seebeck coefficient, $\sigma$ is the electrical conductivity, $T$ is the absolute temperature, $\kappa_e$ and $\kappa_L$ are the electronic and lattice contributions to the total thermal conductivity $\kappa$, respectively. Obviously, obtaining a high $ZT$ requires a high $S$, a high $\sigma$ and a low $\kappa$ simultaneously. But it is a great challenge because of the strong coupling of these parameters.[3,4] The most commonly pursued approaches to enable high $ZT$ include the manipulation of the density of states (DOS) by band engineering for high power factor ($S^2\sigma$),[5] such as doping,[6-8] introducing resonant states in the vicinity of the Fermi level,[9,10] and achieving band convergence/degeneracy,[11-19] *etc*. Especially, band engineering to converge multiple band valleys has been demonstrated to be a robust strategy for achieving a high power factor,[11] and thus it is good for TE materials.[20]

Temperature is one of the intrinsic factors for affecting band convergence and further the TE properties. For instance, in PbX (X = S, Se, and Te), the change of band gap is positively correlated with temperature, and the convergence of the valence band maximum (VBM) at **L** and **Σ** occurs at a high temperature.[21] For *n*-type CoSb$_3$ skutterudites, a secondary conduction band with 12 conducting carrier pockets (which converges with the primary band at high temperatures) is responsible for its extraordinary TE performance.[22] In principle, the temperature effect on the band structure is related to two major mechanisms:[23] the lattice thermal expansion, and the lattice dynamics, such as phonon-induced atomic vibrations. Generally, the latter is defined as the electron-phonon renormalization (EPR).[24-27] So far, the temperature-dependent band gap in lots of semiconductors has been verified in experiments and theoretical calculations.[23,28-39] Recently, we have developed the state-of-art EPR method, and applied it to study the temperature effect on the band structure in traditional semiconductors,[23] TE skutterudites[34,35] and Mg$_2$Si$_{1-x}$Sn$_x$,[36] as well as optoelectronic perovskites[37,38] and pyrite FeS$_2$.[39] Especially for MAPbI$_3$ perovskite, the temperature-induced band dispersion (that is, carrier effective mass) renormalization and further its effect on the charge mobility have been firstly discussed.[37] However, the design and screening of functional materials based on the temperature-dependent band structure has been barely as yet. For TE materials, there are two questions: (1) how does the temperature affect the band convergence and what is its nature; (2) can we efficiently screen some materials with temperature-induced band convergence and so that excellent TE properties.

In order to answer the questions above, the half-Heusler (HH) compounds are studied as the examples in present work. Due to the good TE properties, HHs have emerged as the potential TE materials over the past few decades.[40-42] However, the temperature-dependent band structure, and the possible temperature-induced band convergence which is beneficial to enhance TE properties, have never been fully explored. Here, TiCoSb and NbFeSb are firstly selected as the examples, and we apply the multiple methods[23,26] of EPR to reveal the different temperature responses on the electronic structure properties of the 1$^{st}$ and 2$^{nd}$ conduction band minimum (CBM1, CBM2) in two systems. The two key factors for conduction band convergence at a given temperature for HHs ($ABX$ type) are further found out: (1) $A$ and $B$ sites elements both contribute to the CBM2, while CBM1 mainly comes from $A$; (2) the energy difference between CBM2 and CBM1 should be small enough by the calculation without temperature effect. Using these as the criteria, 3 HHs, namely TiRhBi, TiPtSn and NbPtTl, are screened out from our *MatHub-3d* database.[43,44] They all have the conduction band convergence at mid-high temperature by the further EPR calculations. Taking TiPtSn as the example, the predicted Seebeck coefficient is obviously improved at mid-high temperature. Finally, our experimental transport measurement of TiPtSn verifies the theoretical results. Our work is significant as considering the EPR effect, which is a full understanding of the electronic structure and an accurate prediction of electrical transport at a given temperature. We provide a clearer route forward to the band engineering in the HHs for further improvement of *ZT*.

**Result and discussions**

**Proposal of the band convergence screening criteria.** HH is a ternary intermetallic compound, which is usually expressed by the chemical formula *ABX* with the space group of F$\bar{4}$3m. For TE HHs, *A* is commonly an early transition (IIIB or IVB) or rare-earth metal, *X* is a main group element (IVA or VA), and *B* is normally a transition metal between *A* and *X* in the Periodic Table.[40] The high symmetry of this kind of material makes it having a large band degeneracy and a large DOS. Therefore, it often has the high electrical transport properties and becoming a kind of TE material with full development prospects.[41,42] Among the typical HHs, TiCoSb and NbFeSb are selected in this work to study the temperature-dependent band structure.

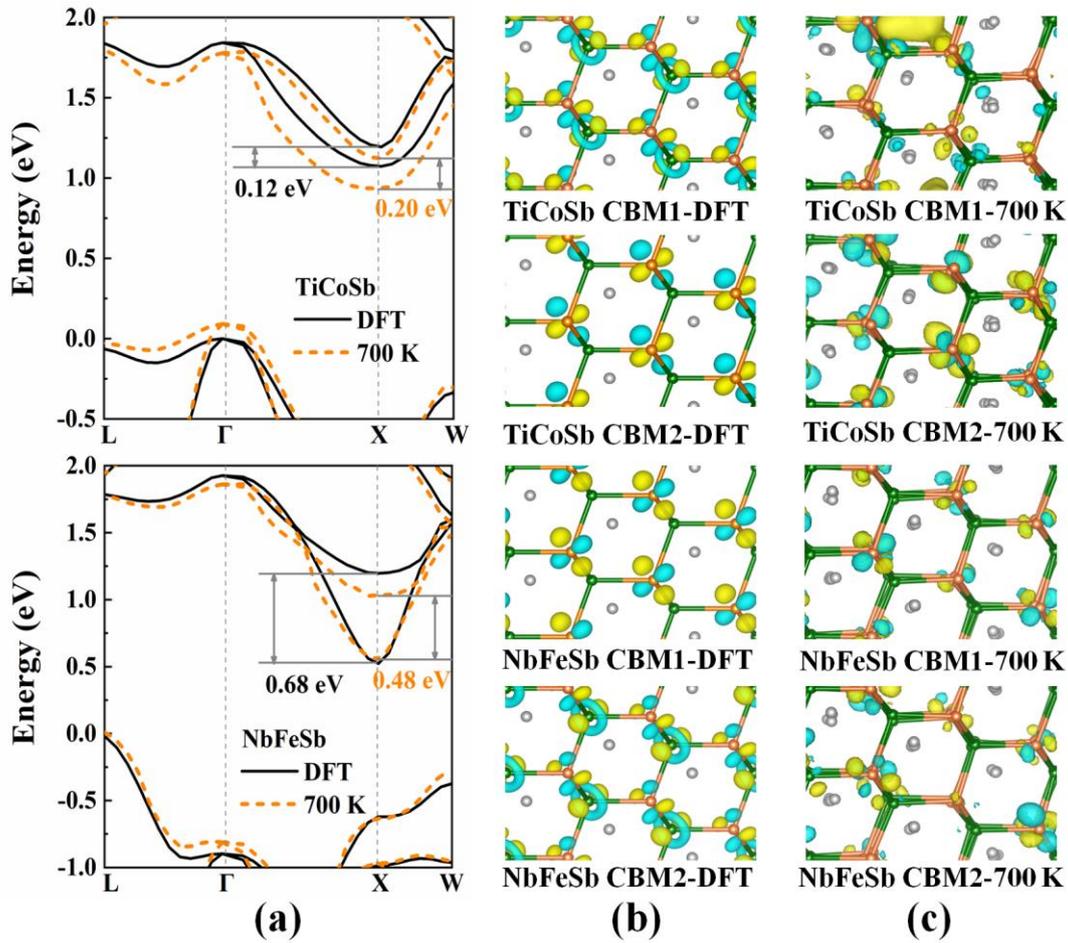

**Figure 1.** TiCoSb and NbFeSb: (a) Band structure without temperature effect (black-solid line, defined as DFT) and that at 700 K (orange-dashed line). The temperature effect includes the lattice expansion and vibration. The energy difference between CBM1 and CBM2 is shown. The high-symmetry **k**-points are **L** (0.5 0.5 0.5), **Γ** (0 0 0), **X** (0.5 0 0.5), **W** (0.5 0.25 0.75). (b) and (c) are the wavefunctions of CBM1 and CBM2 by DFT and at 700 K, respectively. The orange, green and grey ball is Ti/Nb, Co/Fe, and Sb, respectively.

By using the quasi-harmonic approximation (QHA),[45] the lattice parameters of TiCoSb and NbFeSb at different temperatures are determined (Figure S1 in the Supporting Information). Then, the band structures of TiCoSb and NbFeSb at different temperatures are calculated by using the Allen-Heine-Cardona (AHC) theory[26] (Figures S2 and S3), and the cases at 700 K as well as without temperature effect (defined as DFT) are shown in Figure 1a. As increasing to 700 K, the energy difference $E_c$ between CBM1 and CBM2 at **X** point increases from 0.12 eV to 0.20 eV for TiCoSb, while it decreases from 0.68 eV to 0.48 eV for NbFeSb. Although the so large $E_c$ (comparing with $2k_BT$) indicates no band convergence in two HHs, the opposite phenomenon is drawn our attention. As analyzed in our previous works,[34-39] the temperature-induced band gap change or band convergence is due to the different responses of the band energy levels to the temperature. More fundamentally, it originates from the destruction of chemical bonds caused by the temperature-induced lattice expansion and vibration. Thus, we firstly examine the chemical characteristics of CBM1 and CBM2 in TiCoSb and NbFeSb by taking advantages of the wavefunction at CBM (Figure 1b), the projected band structure (Figure S4a) and band-resolved projected crystal orbital Hamilton (pCOHP, Figure S4b). It is found that the CBM1 of TiCoSb and CBM2 of NbFeSb mainly come from the contributions of *A* (Ti/Nb) and *B* (Co/Fe) elements, showing the antibonding between *A* and *B*. Differently, only *A* element Ti or Nb contributes to the CBM2 of TiCoSb or CBM1 of NbFeSb. As mentioned before,[37,38] the antibonding is unstable and its energy would decrease by considering the temperature-induced bond change. Further understanding the temperature effect, the wavefunction of structure at 700 K is shown in Figure 1c. To obtain the structure as including the EPR effect at 700 K, the one-shot method is used.[23] Interestingly, the morphology and distribution of wavefunctions for CBM2 of TiCoSb and CBM1 of NbFeSb at 700 K are almost the same to these without temperature effect (Figure 1b). However, these for CBM1 of TiCoSb and CBM2 of NbFeSb are affected more severely, showing a larger electronic structural disorder. Therefore, the antibonding states are broken, and the energy level of CBM1 of TiCoSb (CBM2 of NbFeSb) drops down with temperature, while that of CBM2 of TiCoSb (CBM1 of

NbFeSb) has relatively small changes (Figure S5), and resulting to the increase (decrease) of $E_c$. The case in NbFeSb is purposeful, however, there is no band convergence occurred even at high temperature, which should be due to the large $E_c$ even that without temperature effect (0.68 eV).

The band renormalization by temperature in TiCoSb and NbFeSb gives us the inspiration. That is, the HHs may have the conduction band convergence on a **k**-point band edge at a given temperature if (1) the antibonding of *A* and *B* elements mainly contributes to the CBM2, and CBM1 only comes from the contribution of *A* element; (2) the energy difference between CBM1 and CBM2 by the calculation without temperature effect is small. In the next section, we will screen the HHs following these criteria.

**Screening HHs with band convergence.** Figure 2 is the workflow of HHs screening. In order to obtain the HHs with semiconductor characteristic and structural stability, the band gap $E_g > 0.1$ eV and phonon frequency $\omega > 0$ are added as the first two screening criteria, and we get 109 out of the 274 HHs from the *MatHub-3d* database. Further considering the screening criteria for band convergence as discussed above, 3 HHs are stood out, which are TiRhBi, TiPtSn and NbPtTl. Specifically, $10k_BT$ at 300 K (~0.26 eV) is used as the threshold value of $E_c$ for the calculation without temperature effect. The corresponding chemical characteristic analysis of three systems are shown in Figure S6. Then, the band structures at different temperatures for the three HHs are calculated, and they all exhibit band convergence at mid-high temperature indeed (Figures S7-S10). As the example, the electrical transport prediction and corresponding experimental measurement in TiPtSn are finally carried out to show the effect of temperature-induced band convergence on TE properties. The details of TiPtSn are discussed in the next section. In addition, 109 HHs' band structures at different temperatures are calculated, and will be stored in *MatHub-3d* for the future study.

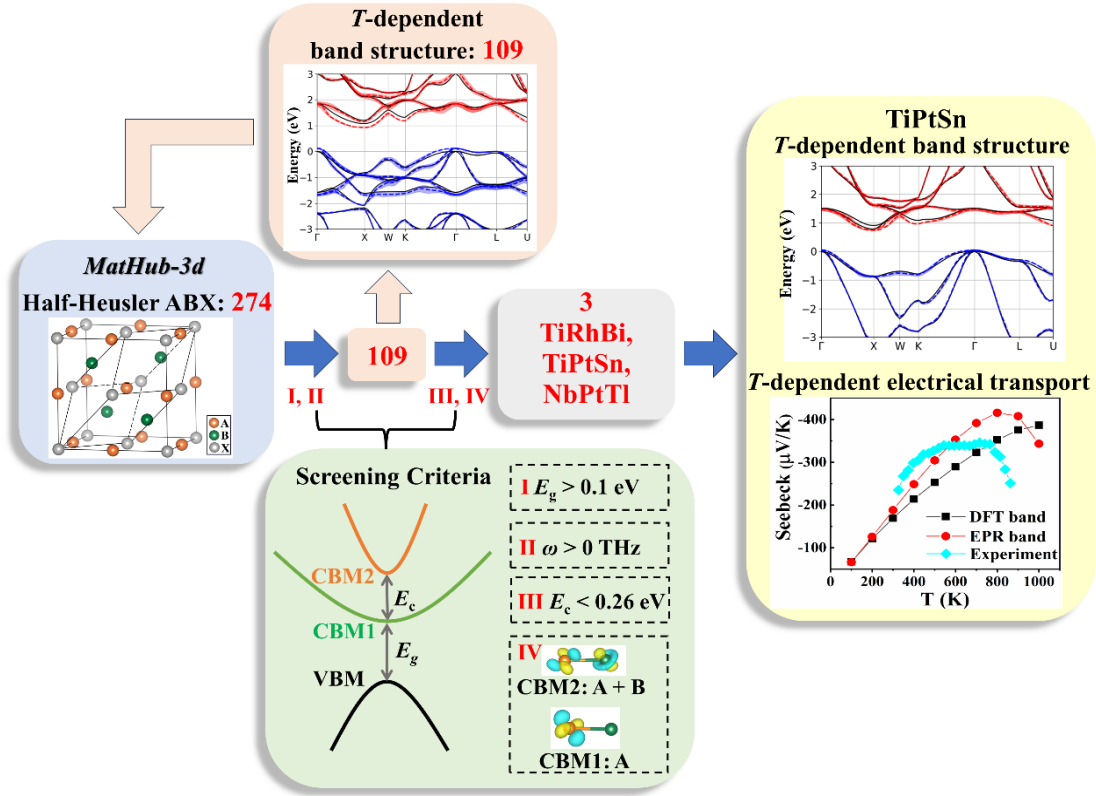

**Figure 2.** Workflow of half-Heuslers screening in present work.

**Temperature-dependent band structure and electrical transport properties in TiPtSn.** The band structures of TiPtSn at 300 K and 700 K are shown in Figure 3a. The case without temperature effect is also plotted for comparing. As expected, the energy of CBM1 and CBM2 at **X** point is obviously overlapped at 700 K. In order to further explore the temperature range of band convergence, the energy difference $E_c$ between CBM1 and CBM2 as the function of temperature is examined (Figure 3b). With the increase of temperature, $E_c$ decreases gradually and then increases. The effective band convergence occurs at ~ 340 K and above, where $E_c < 2k_B T$. Because the energy level of CBM2 is even lower than that of CBM1 at high temperature (Figure S8), $E_c$ increases at > 700 K. The corresponding results for TiRhBi and NbPtTl are shown in Figures S11 and S12, which have the similar temperature law of $E_c$.

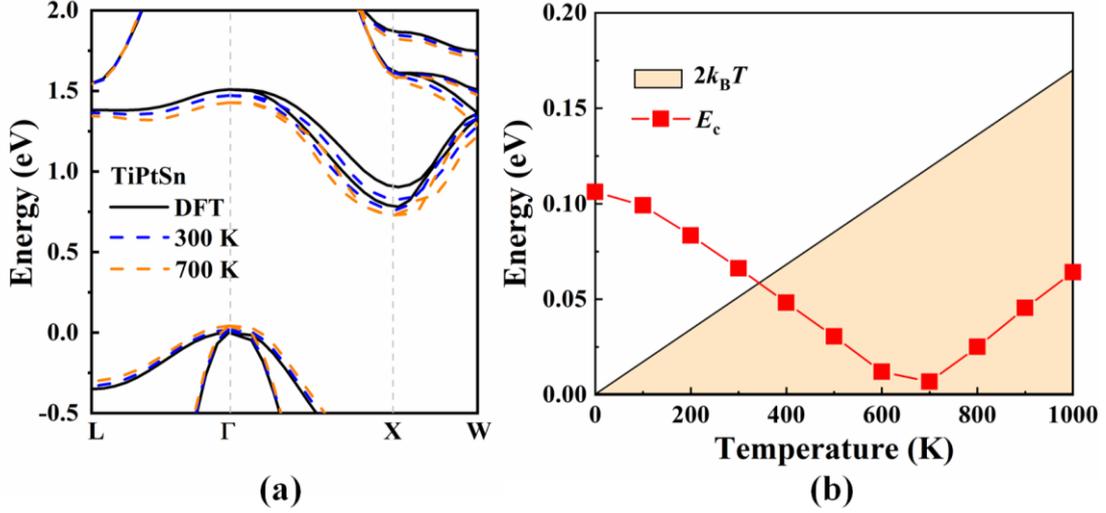

**Figure 3.** (a) Band structure of TiPtSn without temperature effect (DFT, black line), at 300 K (blue-dashed line), and at 700 K (orange-dashed line), respectively. (b) Temperature dependence of the energy difference $E_c$ between CBM1 and CBM2 for TiPtSn. The black line represents a Fermi distribution with $2k_BT$.

In order to confirm the improvement of TE properties by band convergence, the electrical transport properties in TiPtSn are carried out based on the Boltzmann transport theory and relaxation time approximation.[46] To simplify the calculation of relaxation time, the constant electron-phonon coupling such as deformation potential theory is used.[47-50] Notably, it is our first time that using the renormalized electron energy and velocity to predict the transport parameters. It is updated in our *TransOpt* code[51] and the details can be found in the section of Method. Comparing with other electrical transport parameters, the Seebeck coefficient $S$ is highly dependent on the band structure and less dependent on the relaxation time, thus we mainly focus on it. Figures 4a exhibits the predicted $S$ as the function of electron concentration at 700 K. Due to the degenerate bands and thus larger DOS near the Fermi level, the $S$ with 700 K band is larger than that with 300 K as well as without temperature effect. Further using the $S$ and the single parabolic band model,[52] the predicted effective mass enlarges as considering the temperature effect. The similar behavior is found for transport at 300 K (Figure S13). Especially, the effective mass increases from 2.79 $m_0$ at 300 K to 6.49 $m_0$ at 700 K ($m_0$ is the bare electron mass). Meanwhile, the temperature-dependent $S$ at different electron concentration (Figure 4b) is examined. The temperature-induced band convergence makes $S$ enlarged at mid-high temperature, *e.g.*, at 600 K with $10^{20}$ cm$^{-3}$, the $S$ with and without renormalized band is 352.83 µV/K and 289.52 µV/K, respectively. In particular, due to the band gap reduction with temperature (Figure S9), there is a significant bipolar effect near ~800 K with $10^{20}$ cm$^{-3}$ as considering the renormalized band (red-solid symbol in Figure 4b). Therefore, the temperature effect plays an important role on HH's band structure and even electrical transport, and it should not be neglected in the theoretical calculation.

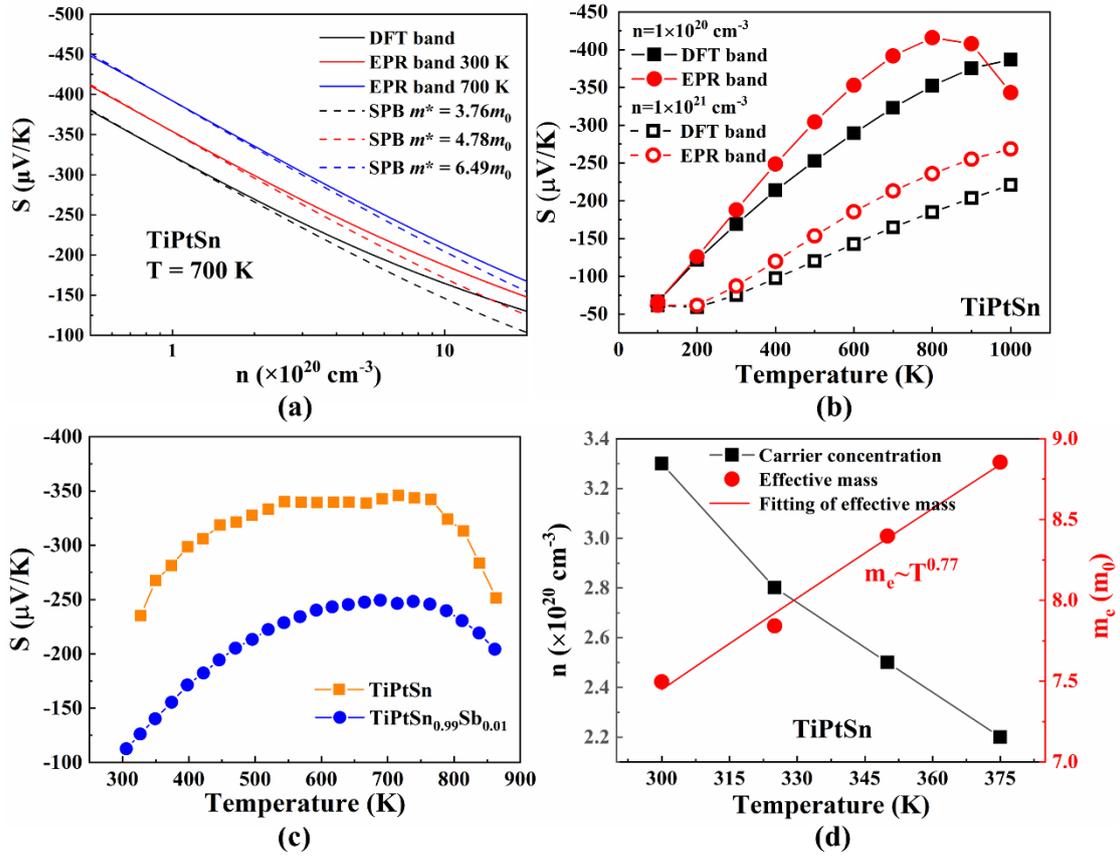

**Figure 4.** (a) Calculated Seebeck coefficient $S$ at 700 K as the function of electron concentration in TiPtSn. Black, red and blue lines represent the $S$ obtained by using the band without temperature effect, at 300 K, and at 700 K, respectively. The corresponding effective mass obtained by the single parabolic band (SPB) model is also shown in dashed line. (b) Temperature-dependent Seebeck coefficient $S$ in TiPtSn with electron concentration $1 \times 10^{20}$ cm$^{-3}$ (solid symbol) and $1 \times 10^{21}$ cm$^{-3}$ (hollow symbol), respectively. The black / red data represent the $S$ obtained by using the band without / with temperature effect. (c) The measured Seebeck coefficient $S$ as the function of temperature in TiPtSn and 1% Sb doped TiPtSn$_{0.99}$Sb$_{0.01}$, respectively. (d) The experimental electron concentration and corresponding effective mass as the function of temperature in TiPtSn, respectively, the temperature dependence of effective mass is also shown.

To further verify the theoretical screening and prediction, the samples of TiPtSn and its 1% Sb doping TiPtSn$_{0.99}$Sb$_{0.01}$ are synthetized (X-ray diffraction patterns in Figure S14a), and the electrical transport properties are then measured. As shown in Figure 4c, the experimental $S$ of TiPtSn increases from 300 K to 500 K, afterwards, it is almost unchanged until obviously decreasing > 750 K, resulting in a bipolar effect. The $S$ of doping 1% Sb TiPtSn$_{0.99}$Sb$_{0.01}$ with the higher electron concentration has the similar temperature law, while the value becomes smaller. Because of the larger electrical conductivity in TiPtSn$_{0.99}$Sb$_{0.01}$ (Figure S14b), its power factor maximum is somehow larger than that of TiPtSn (Figure S14d). Besides the electrical transport properties, the thermal part is also measured, and finally the $ZT$ is predicted (Figure S15). Although the power factor and further the $ZT$ in TiPtSn systems are small as comparing to other outstanding TE materials, the purpose of our experiment is to illustrate the temperature-induced band convergence and the enhanced electrical transport properties caused by it. It is worth noting that the predicted $S$ with renormalized band is closer to the experiment one especially at mid temperature. For example: the measured $S$ at ~ 370 K is 281.56 μV/K with 2.2 × 10$^{20}$ cm$^{-3}$ in TiPtSn (Figure 4c and 4d) and 155.33 μV/K with 1.2 × 10$^{21}$ cm$^{-3}$ in TiPtSn$_{0.99}$Sb$_{0.01}$ (Figures 4c and S16), respectively; the calculated values at the same temperature are ~ 229.80 μV/K with 1 × 10$^{20}$ cm$^{-3}$ and ~ 110.76 μV/K with 1 × 10$^{21}$ cm$^{-3}$. Furthermore, the temperature dependence of the effective mass $m^*$ in TiPtSn and TiPtSn$_{0.99}$Sb$_{0.01}$ are deduced from the measurements of the electrical transport (Figures 4d and S16). Especially for TiPtSn, the $m^*$ enlarges with temperature, indicating the increase of DOS near the Fermi level, which is also demonstrated by above calculation. Therefore, the experiment confirms the theoretical prediction, that is, the increase of temperature causes the conduction band convergence in TiPtSn, and further the enlarged DOS and Seebeck.

Worthy of note is that the $m^*$ in TiPtSn follows ~ $T^{0.77}$, and the temperature dependence of $m^*$ in TE materials is widely known for a long time, such as in lead chalcogenides with ~ $T^{0.4}$.[53,54] However, there is no in-depth theoretical explanation or full understanding. As performing the EPR calculation, our work is the first time that revealing the origin of temperature dependence of $m^*$, and more detailed studies in more TE materials will be carried out in the future.

**Conclusions**

In this work, we propose a new strategy that is optimizing TE performance through screening HH compounds with possible temperature-induced band convergence. Taking the TiCoSb and NbFeSb as the examples, two key factors for conduction band convergence at a given temperature are revealed. Using these as the criteria, TiRhBi, TiPtSn, and NbPtTl are then screened out from 274 HHs in *MatHub3d*, and they all show the conduction band convergence at mid-high temperature. Further the electrical transport calculation definitely shows the enlarged Seebeck coefficient in TiPtSn, which is due to the temperature-induced band convergence. The finally experimental measurement verifies the theoretical results. Hence, the temperature plays an important role on the band structure and further the electrical transport, and cannot be neglected in theoretical prediction. Our work has the deep understanding of temperature-dependent electronic structures, and screen out the HHs with band convergence and fine electrical transport properties. It can provide a clearer route forward to band structure engineering as well as the extension to more studies of the basic TE properties in the HHs.

**Methods**

**Computational model and parameter setting.** The band structure is influenced by the temperature effect from two aspects: the lattice thermal expansion and phonon vibration.[23] In this study, the QHA[45] method is used to determine the lattice constant at a given temperature $T$, which is implemented by using the Vienna *ab initio* Simulation Package (VASP)[55,56] and Phonopy code.[57] Wherein, Perdew-Burke-Ernzerhof (PBE)[58] type exchange-correlation functional is used, the projector augmented wave method[59] is applied with the plane-wave cutoff energy of 520 eV, and the $4 \times 4 \times 4$ supercell is used for phonon calculation. Therefore, the structure with lattice parameters at $T$ is obtained (Figures S1 and S7). Then, in order to consider the contribution from phonon vibration, the many-body perturbation approach: AHC theory is employed as implemented in ABINIT package.[60] In the framework of AHC theory, the electron-phonon coupling is evaluated by the density functional perturbation theory (DFPT).[61] The electron-nucleus interactions are described by using the norm-conserving pseudopotentials[62] with the plane-wave cut off of 35 Hartree, **k**-mesh of 6 × 6 × 6 and **q**-mesh of 10 × 10 × 10. After the DFPT calculation at PBE level, the electron self-energy $\Sigma_{n\mathbf{k}}^{ep}(T)$ for the $n^{\text{th}}$ band and the wave-vector **k** at $T$ can be determined by the AHC theory[26]

$$\Sigma_{n\mathbf{k}}^{ep}(T) = \Sigma_{n\mathbf{k}}^{\text{FM}}(T) + \Sigma_{n\mathbf{k}}^{\text{DW}}(T), \tag{1}$$

here the Fan-Migdal (FM) part under the quasi-static approximation and the static Debye-Waller (DW) part are respectively given by

$$\Sigma_{n\mathbf{k}}^{\text{FM}}(T) = \sum_{m,v} \int \frac{d\mathbf{q}}{\Omega_{BZ}} |g_{mnv}(\mathbf{k},\mathbf{q})|^2 \left[ \frac{n_{\mathbf{q}v}(T) + f_{m\mathbf{k}+\mathbf{q}}(\varepsilon_F,T)}{\varepsilon_{n\mathbf{k}} - \varepsilon_{m\mathbf{k}+\mathbf{q}} + \omega_{\mathbf{q}v} + i\eta} + \frac{n_{\mathbf{q}v}(T) + 1 - f_{m\mathbf{k}+\mathbf{q}}(\varepsilon_F,T)}{\varepsilon_{n\mathbf{k}} - \varepsilon_{m\mathbf{k}+\mathbf{q}} - \omega_{\mathbf{q}v} + i\eta} \right], \tag{2}$$

$$\Sigma_{n\mathbf{k}}^{\text{DW}}(T) = \sum_{m,v} \int \frac{d\mathbf{q}}{\Omega_{BZ}} [2n_{\mathbf{q}v}(T) + 1] \frac{g_{mnv}^{2,DW}(\mathbf{k},\mathbf{q})}{\varepsilon_{n\mathbf{k}} - \varepsilon_{m\mathbf{k}+\mathbf{q}}}. \tag{3}$$

Where $f_{m\mathbf{k}+\mathbf{q}}(\varepsilon_F, T)$ and $n_{\mathbf{q}v}(T)$ are the Fermi-Dirac and Bose-Einstein occupation functions with $T$ and the Fermi level $\varepsilon_F$. The integration is performed over the **q**-points in the Brillouin zone (BZ) of volume $\Omega_{BZ}$, $\omega_{\mathbf{q}v}$ is the phonon frequency on the branch $v$ and wave-vector **q**, $\eta$ is a positive real infinitesimal. $g_{mnv}^{2,DW}(\mathbf{k},\mathbf{q})$ is an effective matrix element that, within the rigid-ion approximation, can be expressed in terms of the standard first-order $g_{mnv}(\mathbf{k},\mathbf{q})$ matrix elements by exploiting the invariance of the quasi-particle energies under infinitesimal translation. The renormalized electron energy $\tilde{\varepsilon}_{n\mathbf{k}}$ at $T$ is the sum of the bare Kohn-Sham eigenvalue $\varepsilon_{n\mathbf{k}}$ and the real part of $\Sigma_{n\mathbf{k}}^{ep}(T)$ as following:

$$\tilde{\varepsilon}_{n\mathbf{k}} = \varepsilon_{n\mathbf{k}} + \Re\Sigma_{n\mathbf{k}}^{ep}(T). \tag{4}$$

By using the $\tilde{\varepsilon}_{n\mathbf{k}}$, the electrical transport parameters as considering the electron energy renormalization can be determined under the Boltzmann transport theory as implemented in *TransOpt*[51] code. Specifically, electrical conductivity $\sigma$ and Seebeck coefficient *S* can be calculated as following:[50,51]

$$\sigma_{\alpha\beta}(\varepsilon_F, T) = \sum_n \int \frac{d\mathbf{k}}{\Omega_{BZ}} \tilde{v}_{n\mathbf{k}\alpha} \tilde{v}_{n\mathbf{k}\beta} \tilde{\tau}_{n\mathbf{k}} \left[ -\frac{\partial \tilde{f}_{n\mathbf{k}}(\varepsilon_F,T)}{\partial \tilde{\varepsilon}_{n\mathbf{k}}} \right], \tag{5}$$

$$S_{\alpha\beta}(\varepsilon_F, T) = \frac{1}{eT} \sigma_{\alpha\beta}(\varepsilon_F, T)^{-1} \sum_n \int \frac{d\mathbf{k}}{\Omega_{BZ}} \tilde{v}_{n\mathbf{k}\alpha} \tilde{v}_{n\mathbf{k}\beta} \tilde{\tau}_{n\mathbf{k}} (\varepsilon_F - \tilde{\varepsilon}_{n\mathbf{k}}) \left[-\frac{\partial \tilde{f}_{n\mathbf{k}}(\varepsilon_F, T)}{\partial \tilde{\varepsilon}_{n\mathbf{k}}}\right], \quad (6)$$

where the electron group velocity $\tilde{\mathbf{v}}_{n\mathbf{k}}$, the electron relaxation time $\tilde{\tau}_{n\mathbf{k}}$, and the Fermi-Dirac occupation function $\tilde{f}_{n\mathbf{k}}$ are obtained as replacing the $\varepsilon_{n\mathbf{k}}$ with $\tilde{\varepsilon}_{n\mathbf{k}}$. Especially, $\tilde{\tau}_{n\mathbf{k}}$ is calculated by using the deformation potential approximation, which is written as[49-51]

$$\frac{1}{\tilde{\tau}_{n\mathbf{k}}} = \frac{2\pi k_B T D_{def}^2}{\hbar C} \sum_m \int \frac{d\mathbf{k}'}{\Omega_{BZ}} \delta(\tilde{\varepsilon}_{n\mathbf{k}} - \tilde{\varepsilon}_{m\mathbf{k}'}), \quad (7)$$

here $k_B$ is the Boltzmann constant, $\hbar$ is the reduced Planck constant, $D_{def}$ is the deformation potential of the band edge state, and $C$ is the Young's modulus. According to $\Delta E_{band\,edge} = D_{def} \frac{\Delta\Omega}{\Omega}$ and $\frac{\Delta E_{tot}}{\Omega} = \frac{1}{2} C \left(\frac{\Delta\Omega}{\Omega}\right)^2$ ($\Delta E_{tot}$ is the change in total energy, $\Delta E_{band\,edge}$ is the change in absolute energy level of the band edge, and $\frac{\Delta\Omega}{\Omega}$ is the relative change in the lattice constant at $T$),[48] $D_{def}$ and $C$ can be determined by linear fitting and parabolic fitting at each $T$ with corresponding lattice constant, respectively, and the values at each $T$ are summarized in Table S1. To satisfy the convergence of integrals, the **k**-mesh used in the calculations of the transport properties is 40 × 40× 40. The n-type doping in TiPtSn with different electron concentrations are considered by the rigid shift of the Fermi level.

In order to examine the temperature effect on the wavefunction as shown in Figures 1c and S6, the one-shot method is used for obtaining the effective structure at $T$. Specifically, the vibration-induced atomic displacement $\Delta\tau_{\kappa\alpha}$ ($\kappa$ and $\alpha$ indicate the atom and the Cartesian direction, respectively) at $T$ is determined as following:[23,63]

$$\Delta\tau_{\kappa\alpha} = (M_p/M_\kappa)^{\frac{1}{2}} \sum_v (-1)^{v-1} e_{\kappa\alpha,v} \sigma_{v,T}, \quad (8)$$

here, $M_p$ is the proton mass, $M_\kappa$ is the $\kappa$th nucleus's mass, and the Gaussian width $\sigma_{v,T}$ for the $v^{th}$ normal mode at $T$ is given by

$$\sigma_{v,T}^2 = (\hbar/2M_p \omega_v)(2n_{v,T} + 1), \quad (9)$$

where $n_{v,T} = [exp(\hbar\omega_v/k_B T) - 1]^{-1}$ is the Bose-Einstein distribution. It considers the contributions from all phonon vibrations to the structure, and the details can be found in our previous work.[23] Besides, the band-resolved pCOHP in Figures S4 and S5 is implemented using the LOBSTER code.[64]

**Experimental sample preparation and TE properties measurement.** *Sample Synthesis:* Polycrystalline samples TiPtSn$_{1-x}$Sb$x$ ($x$ = 0, 0.01) are synthesized under an argon atmosphere by arc melting stoichiometric amounts of high-purity elements Ti (granules, 99.98%), Pt (filaments, 99.999%), and Sn (granules, 99.99%). The ingots are remelted 3 times to ensure compositional homogeneity. Additional 3 wt% Sn is added to compensate for the evaporation loss during arc melting. Then the arc melted ingot is crushed and ball-milled for 5 hours by a FRITSCH Pulverisette-7 Premium Line ball-milling machine. The as-milled fine ground powders are then loaded into a graphite die with an inner diameter of 12.7 mm and compacted into dense pellets by spark plasma sintering (LABOX-325GH-C1, Japan) at 1223 K for 15 min under an axial pressure of 50 MPa in vacuum. The as-sintered samples with relative densities over 98% are cut into different shapes for measuring the thermal diffusivity (Φ10 mm × 1 mm), Hall effect (0.6 mm × 4 mm × 8 mm), Seebeck coefficient, and electrical conductivity (2 mm × 3 mm × 12 mm).

*Sample characterization*: X-ray diffraction (XRD) patterns of all samples are obtained by the Cu-Kα (λ = 1.54185 Å) radiation (Rigaku SmartLab, Japan) at room temperature. The Seebeck coefficient and electrical conductivity are measured in helium atmosphere utilizing a commercially available instrument (ZEM-3, ULVAC-RIKO, Japan). The carrier concentration and mobility are measured by a comprehensive physical property measuring system (PPMS, Quantum Design, USA). and the Hall carrier concentration $n_H$ is calculated according to $n_H = 1/(eR_H)$. The calculated DOS effective mass $m^*$ is by the single parabolic band model. This is using the equation[52]

$$S = \frac{8\pi^2}{3eh^2} m^* \left(\frac{\pi}{3n_H}\right)^{2/3}. \tag{10}$$

**Acknowledgements**

This work was supported by the National Key Research and Development Program of China (no. 2021YFB3502200), and the National Natural Science Foundation of China (grant no. 52172216, 92163212, and 52302282). Numerical computations were performed on Hefei Advanced Computing Center and Shanghai Technical Service

Center of Science and Engineering Computing, Shanghai University.

**References**

[1] Yang J, Yip H-L, Jen A K Y. Rational design of advanced thermoelectric materials. Advanced Energy Materials, 2013, 3(5): 549-565.

[2] Snyder G J, Toberer E S. Complex thermoelectric materials. Nature Materials, 2008, 7(2): 105-114.

[3] Wei T-R, Tan G, Zhang X, et al. Distinct impact of alkali-ion doping on electrical transport properties of thermoelectric p-type polycrystalline SnSe. Journal of the American Chemical Society, 2016, 138(28): 8875-8882.

[4] Dong J, Wu C-F, Pei J, et al. Lead-free MnTe mid-temperature thermoelectric materials: facile synthesis, p-type doping and transport properties. Journal of Materials Chemistry C, 2018, 6(15): 4265-4272.

[5] Tan G, Zhao L-D, Shi F, et al. High thermoelectric performance of p-type SnTe via a synergistic band engineering and nanostructuring approach. Journal of the American Chemical Society, 2014, 136(19): 7006-7017.

[6] Liang G, Lyu T, Hu L, et al. $(GeTe)_{1-x}(AgSnSe_2)_x$: Strong atomic disorder-induced high thermoelectric performance near the ioffe–regel limit. ACS Applied Materials & Interfaces, 2021, 13(39): 47081-47089.

[7] Zhao L, He J, Wu C-I, et al. Thermoelectrics with Earth abundant elements: high performance p-type PbS nanostructured with SrS and CaS. Journal of the American Chemical Society, 2012, 134(18): 7902-7912.

[8] Ohta M, Biswas K, Lo S H, et al. Enhancement of thermoelectric figure of merit by the insertion of MgTe nanostructures in p-type PbTe doped with $Na_2Te$. Advanced Energy Materials, 2012, 2(9): 1117-1123.

[9] Wu L, Li X, Wang S, et al. Resonant level-induced high thermoelectric response in indium-doped GeTe. NPG Asia Materials, 2017, 9(1): e343-e343.

[10] Zhang Q, Liao B, Lan Y, et al. High thermoelectric performance by resonant dopant indium in nanostructured SnTe. Proceedings of the National Academy of Sciences, 2013, 110(33): 13261-13266.

[11] Pei Y, Shi X, LaLonde A, et al. Convergence of electronic bands for high performance bulk thermoelectrics. Nature, 2011, 473(7345): 66-69.

[12] Zhao L D, Wu H J, Hao S Q, et al. All-scale hierarchical thermoelectrics: MgTe in PbTe facilitates valence band convergence and suppresses bipolar thermal transport for high performance. Energy & Environmental Science, 2013, 6(11): 3346-3355.

[13] Liu W, Tan X, Yin K, et al. Convergence of conduction bands as a means of enhancing thermoelectric performance of *n*-type $Mg_2Si_{1-x}Sn_x$ solid solutions. Physical Review Letters, 2012, 108(16).

[14] Zeier W G, Zhu H, Gibbs Z M, et al. Band convergence in the non-cubic chalcopyrite compounds $Cu_2MGeSe_4$. Journal of Materials Chemistry C, 2014, 2(47): 10189-10194.

[15] Zhang J, Liu R, Cheng N, et al. High-Performance pseudocubic thermoelectric materials from non-cubic chalcopyrite compounds. Advanced Materials, 2014, 26(23): 3848-3853.

[16] Lv H Y, Lu W J, Shao D F, et al. Enhanced thermoelectric performance of phosphorene by strain-induced band convergence. Physical Review B, 2014, 90(8): 085433.

[17] Zheng Z, Su X, Deng R, et al. Rhombohedral to cubic conversion of GeTe via MnTe alloying leads to ultralow thermal conductivity, electronic band convergence, and high thermoelectric performance. Journal of the American Chemical Society, 2018, 140(7): 2673-2686.

[18] Tan G, Hao S, Hanus R C, et al. High Thermoelectric performance in SnTe-$AgSbTe_2$ alloys from lattice softening, giant phonon-vacancy scattering, and valence band convergence. ACS Energy Letters, 2018, 3(3): 705-712.

[19] Wu Y, Chen Z, Nan P, et al. Lattice strain advances thermoelectrics. Joule, 2019, 3(5): 1276-1288.

[20] DiSalvo F J. Thermoelectric cooling and power generation. Science, 1999, 285(5428): 703-706.

[21] Gibbs Z M, Kim H, Wang H, et al. Temperature dependent band gap in PbX (X= S, Se, Te). Applied Physics Letters, 2013, 103(26): 262109.

[22] Tang Y, Gibbs Z M, Agapito L A, et al. Convergence of multi-valley bands as the electronic origin of high thermoelectric performance in $CoSb_3$ skutterudites. Nature Materials, 2015, 14(12): 1223-1228.

[23] Zhang Y, Wang Z, Xi J, et al. Temperature-dependent band gaps in several semiconductors: from the role of electron-phonon renormalization. Journal of Physics-Condensed Matter, 2020, 32(47): 475503.

[24] Poncé S, Gillet Y, Laflamme Janssen J, et al. Temperature dependence of the electronic structure of semiconductors and insulators. The Journal of Chemical Physics, 2015, 143(10): 102813.

[25] Poncé S, Antonius G, Gillet Y, et al. Temperature dependence of electronic eigenenergies in the adiabatic harmonic approximation. Physical Review B, 2014, 90(21): 214304.

[26] Giustino F. Electron-phonon interactions from first principles. Reviews of Modern Physics, 2017, 89(1): 015003.

[27] Giustino F, Louie S G, Cohen M L. Electron-phonon renormalization of the direct band gap of diamond. Physical Review Letters, 2010, 105(26): 265501.

[28] Varshni Y P. Temperature dependence of the energy gap in semiconductors. Physica, 1967, 34(1): 149-154.

[29] Yoshida J, Kita T, Wada O, et al. Temperature dependence of $GaAs_{1-x}Bi_x$ band gap studied by photoreflectance spectroscopy. Japanese Journal of Applied Physics Part 1-Regular Papers Brief Communications & Review Papers, 2003, 42(2A): 371-374.

[30] Vinod E M, Naik R, Faiyas A P A, et al. Temperature dependent optical constants of amorphous $Ge_2Sb_2Te_5$ thin films. Journal of Non-Crystalline Solids, 2010, 356(41-42): 2172-2174.

[31] Quarti C, Mosconi E, Ball J M, et al. Structural and optical properties of methylammonium lead iodide across the tetragonal to cubic phase transition: implications for perovskite solar cells. Energy & Environmental Science, 2016, 9(1): 155-163.

[32] Shang H, Zhao J, Yang J. Assessment of the mass factor for the electron–phonon coupling in solids. The Journal of Physical Chemistry C, 2021, 125(11): 6479-6485.

[33] Park J, Saidi W A, Chorpening B, et al. Applicability of Allen-Heine-Cardona theory on MOX metal oxides and $ABO_3$ perovskites: toward high-temperature optoelectronic applications. Chemistry of Materials, 2022, 34(13): 6108-6115.

[34] Wang Z, Xi J, Ning J, et al. Temperature-dependent band renormalization in $CoSb_3$ skutterudites due to Sb-ring-related vibrations. Chemistry of Materials, 2021, 33(3): 1046-1052.

[35] Ning J, Lei W, Yang J, Xi J. First-principles study of the temperature-induced band renormalization in thermoelectric filled skutterudites. Physical Chemistry Chemical Physics, 2023, 25(38): 26006-26013.

[36] Yao W, Shi B, Hu S, et al. Temperature-induced band gap renormalization in $Mg_2Si$ and $Mg_2Sn$. Physical Review B, 2023, 108(15): 155205.

[37] Xi J, Zheng L, Wang S, et al. Temperature-dependent structural fluctuation and its effect on the electronic structure and charge transport in hybrid perovskite $CH_3NH_3PbI_3$. Journal of Computational Chemistry, 2021, 42(31): 2213-2220.

[38] Ning J, Zheng L, Lei W, et al. Temperature-dependence of the band gap in the all-inorganic perovskite $CsPbI_3$ from room to high temperatures. Physical Chemistry Chemical Physics, 2022, 24(26): 16003-16010.

[39] Li Y, Ning J, Xi J, et al. Weak electron-phonon renormalization effect caused by the counteraction of the different phonon vibration modes in $FeS_2$. Physica Scripta, 2023, 98(6): 065902.

[40] Tritt T M. Recent Trends in Thermoelectric Materials Research II; Academic Press: San Diego, CA, 2001, Chapter 2: 37-72.

[41] Zeier W G, Schmitt J, Hautier G, et al. Engineering half-Heusler thermoelectric materials using Zintl chemistry. Nature Reviews Materials, 2016, 1(6): 16032.

[42] Casper F, Graf T, Chadov S, et al. Half-Heusler compounds: novel materials for energy and spintronic applications. Semiconductor Science and Technology, 2012, 27(6): 063001.

[43] Yao M, Wang Y, Li X, et al. Materials informatics platform with three dimensional structures, workflow and thermoelectric applications. Scientific Data, 2021, 8(1): 236.

[44] http://www.mathub3d.net/.

[45] Togo A, Chaput L, Tanaka I, et al. First-principles phonon calculations of thermal expansion in $Ti_3SiC_2$, $Ti_3AlC_2$, and $Ti_3GeC_2$. Physical Review B, 2010, 81(17): 174301.

[46] Ziman J M. Principles of the Theory of Solids, Cambridge University, London, 1972.

[47] Bardeen J, Shockley W. Deformation potentials and mobilities in non-polar

crystals. Physical Review, 1950, 80(1): 72-80.

[48] Xi J, Long M, Tang L, et al. First-principles prediction of charge mobility in carbon and organic nanomaterials. Nanoscale, 2012, 4(15): 4348-4369.

[49] Xi L, Pan S, Li X, et al. Discovery of high-performance thermoelectric chalcogenides through reliable high-throughput material screening. Journal of the American Chemical Society, 2018, 140(34): 10785-10793.

[50] Xi J, Zhu Z, Xi L, et al. Perspective of the electron-phonon interaction on the electrical transport in thermoelectric/electronic materials. Applied Physics Letters, 2022, 120(19): 190503.

[51] Li X, Zhang Z, Xi J, et al. TransOpt. A code to solve electrical transport properties of semiconductors in constant electron-phonon coupling approximation. Computational Materials Science, 2021, 186: 110074.

[52] Cutler M, Leavy J F, Fitzpatrick, R L. Electronic transport in semimetallic cerium sulfide. Physical Review, 1964, 133(4A): A1143-A1152.

[53] Wang H, Pei Y, LaLonde A D, et al. Weak electron-phonon coupling contributing to high thermoelectric performance in n-type PbSe. Proceedings of the National Academy of Sciences, 2012, 109(25): 9705-9709.

[54] Ravich Y I, Efimova B A, Smirnov I A. Semiconducting lead chalcogenides. 1970, Plenum Press, New York, pp 196-209.

[55] Kresse G, Furthmüller J. Efficient iterative schemes for ab initio total-energy calculations using a plane-wave basis set. Physical Review B, 1996, 54(16): 11169-11186.

[56] Kresse G, Joubert D. From ultrasoft pseudopotentials to the projector augmented-wave method. Physical Review B, 1999, 59(3): 1758-1775.

[57] Togo A, Tanaka I. First principles phonon calculations in materials science, Scripta Materialia, 2015, 108: 1-5.

[58] Perdew J P, Burke K, Ernzerhof M. Generalized gradient approximation made simple, Physical Review Letters, 1997, 78(7): 1396.

[59] Blöchl P E, Projector augmented-wave method, Physical Review B, 1994, 50(24): 17953-17979.

[60] Gonze X, Amadon B, Antonius G, et al. The ABINIT project: Impact, environment and recent development. Computer Physics Communications, 2020, 248: 107042.

[61] Baroni S, de Gironcoli S, Dal Corso A. Phonons and related crystal properties from density-functional perturbation theory. Reviews of Modern Physics, 2001, 73(2): 515-562.

[62] Hamann D R. Optimized norm-conserving Vanderbilt pseudopotentials. Physical Review B, 2013, 88(8): 085117.

[63] Zacharias M, Giustino F. One-shot calculation of temperature-dependent optical spectra and phonon-induced band-gap renormalization. Physical Review B, 2016, 94(7): 075125.

[64] Maintz S, Deringer V L, Tchougréeff A L, et al. LOBSTER: A tool to extract chemical bonding from plane-wave based DFT. Journal of Computational Chemistry, 2016, 37(11): 1030.

# Supporting Information

# for

# Screening of half-Heuslers with temperature-induced band convergence and enhanced thermoelectric properties

Jinyang Xi,[1,†] Zirui Dong,[2,†] Menghan Gao,[1] Jun Luo,[2] and Jiong Yang[1,*]

[1]Materials Genome Institute, Shanghai University, Shanghai 200444, China.

[2]School of Materials Science and Engineering, Shanghai University, Shanghai 200444, China.

[†]The authors contributed equally.

**Corresponding Author:**

[*]Jiong Yang, Email: jiongy@t.shu.edu.cn

**Thermal conductivity measurement.** The thermal diffusivity $\lambda$ of a disk sample with a diameter of 12.7 mm and a thickness of about 1 mm was measured using an automatic pulsed laser thermal conductivity measurement system (LFA457, NETZSCH, Germany). Mass density $\rho$ was measured by the Archimedean method, specific heat capacity ($C_P$) was calculated by Dulonn-Petit law, and total thermal conductivity $\kappa$ was calculated by $\kappa = \rho\lambda C_p$. The lattice thermal conductivity $\kappa_L$ is normally obtained by subtracting the electronic thermal conductivity $\kappa_e$ from the total thermal conductivity $\kappa$ through the Wiedemann–Franz law $\kappa_L = \kappa - \kappa_e = \kappa - L\sigma T$, where $\sigma$, $L$ and $T$ are the electrical conductivity, Lorenz number, and absolute temperature, respectively. However, the bipolar thermal conductivity $\kappa_b$ should be taken into account if the intrinsic excitation presents, which gives the relationship $\kappa_L + \kappa_b = \kappa - L\sigma T$. The Lorenz number is estimated under the framework of single parabolic band model with the assumption of electron-phonon interaction. The employed equations for the Lorenz number can be written as

$$S = \pm \frac{k_B}{e}\left[\frac{(r+\frac{5}{2})F_{r+\frac{3}{2}}(\eta)}{(r+\frac{3}{2})F_{r+\frac{1}{2}}(\eta)} - \eta\right], \tag{S1}$$

$$L = \left(\frac{k_B}{e}\right)^2\left[\frac{(r+\frac{7}{2})F_{r+\frac{5}{2}}(\eta)}{(r+\frac{3}{2})F_{r+\frac{3}{2}}(\eta)} - \left(\frac{(r+\frac{5}{2})F_{r+\frac{3}{2}}(\eta)}{(r+\frac{3}{2})F_{r+\frac{1}{2}}(\eta)}\right)^2\right], \tag{S2}$$

$$F(\xi) = \int_0^\infty \frac{x^i}{1+e^{x-\xi}}dx, \tag{S3}$$

where $S$ is the Seebeck coefficient, $k_B$ the Boltzmann constant, $e$ the free electron charge, $r$ the scattering factor, $F_i$ the $i$-th Fermi integral, $\eta$ the reduced Fermi level (Fermi level over $k_B T$). The scattering factor $r$ is equal to -1/2 assuming that the acoustic phonon scattering dominates the charge transport process.

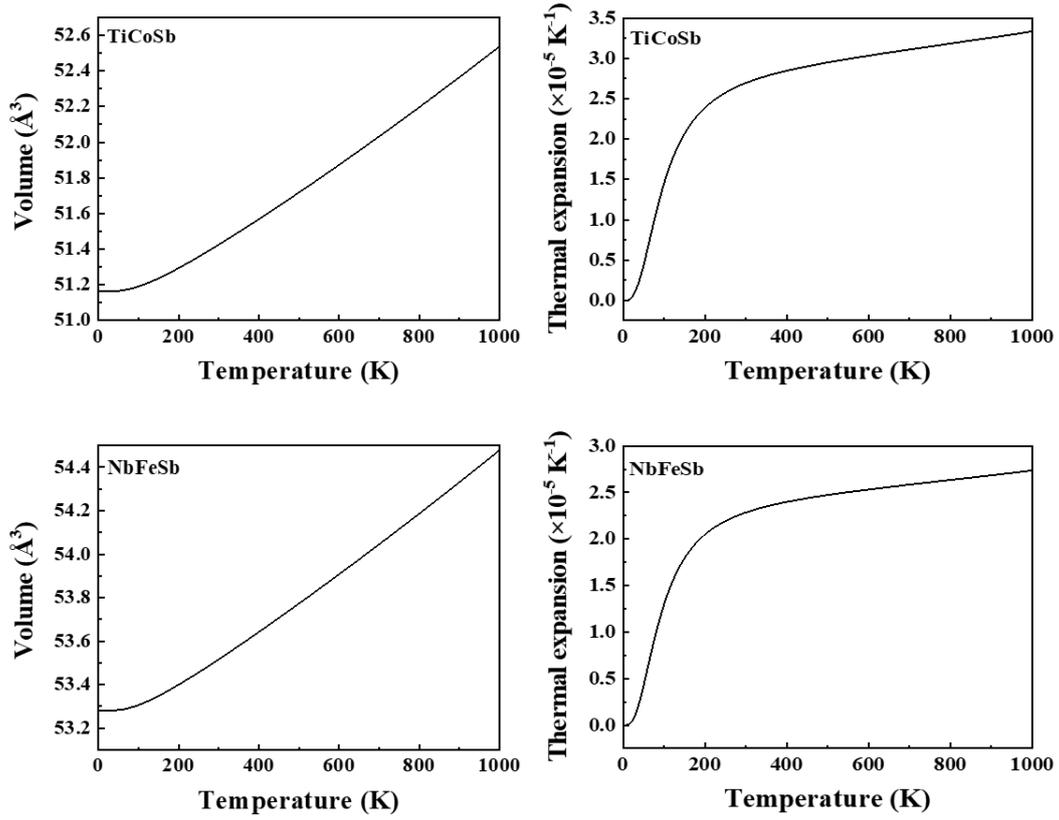

**Figure S1.** Volume-temperature and volumetric thermal expansion coefficient-temperature relations in TiCoSb and NbFeSb, respectively.

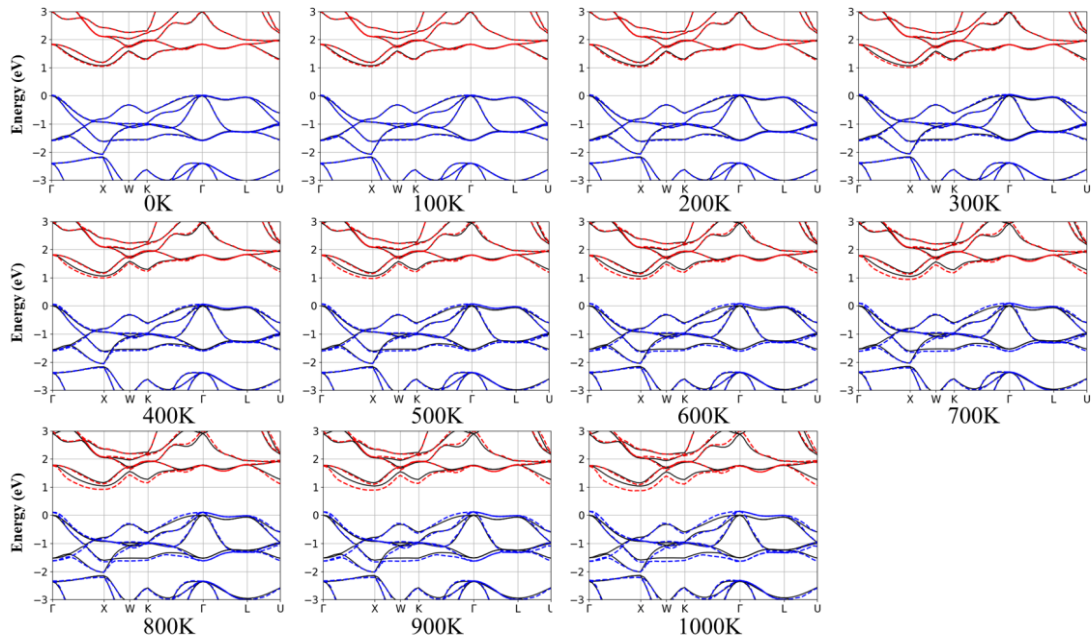

**Figure S2.** Temperature-dependent band structure of TiCoSb from 0 K to 1000 K. The lattice expansion and vibration effects are both considered. The black line is obtained without temperature effect, the blue and red dashed lines are valence and conduction bands at a given

temperature, respectively.

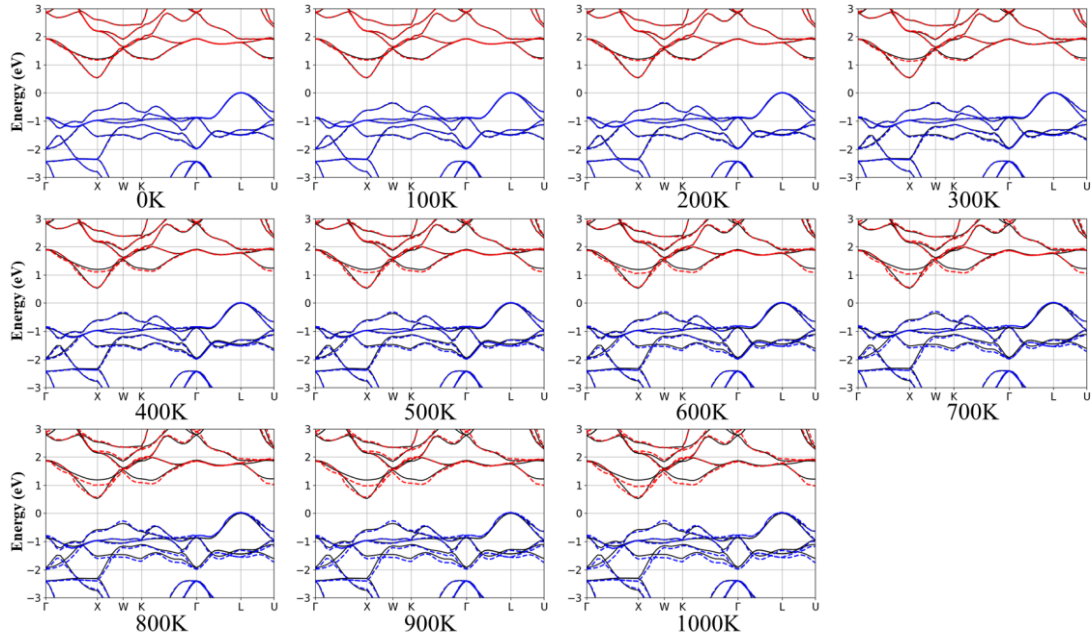

**Figure S3.** Temperature-dependent band structure of NbFeSb from 0 K to 1000 K. The lattice expansion and vibration effects are both considered. The black line is obtained without temperature effect, the blue and red dashed lines are valence and conduction bands at a given temperature, respectively.

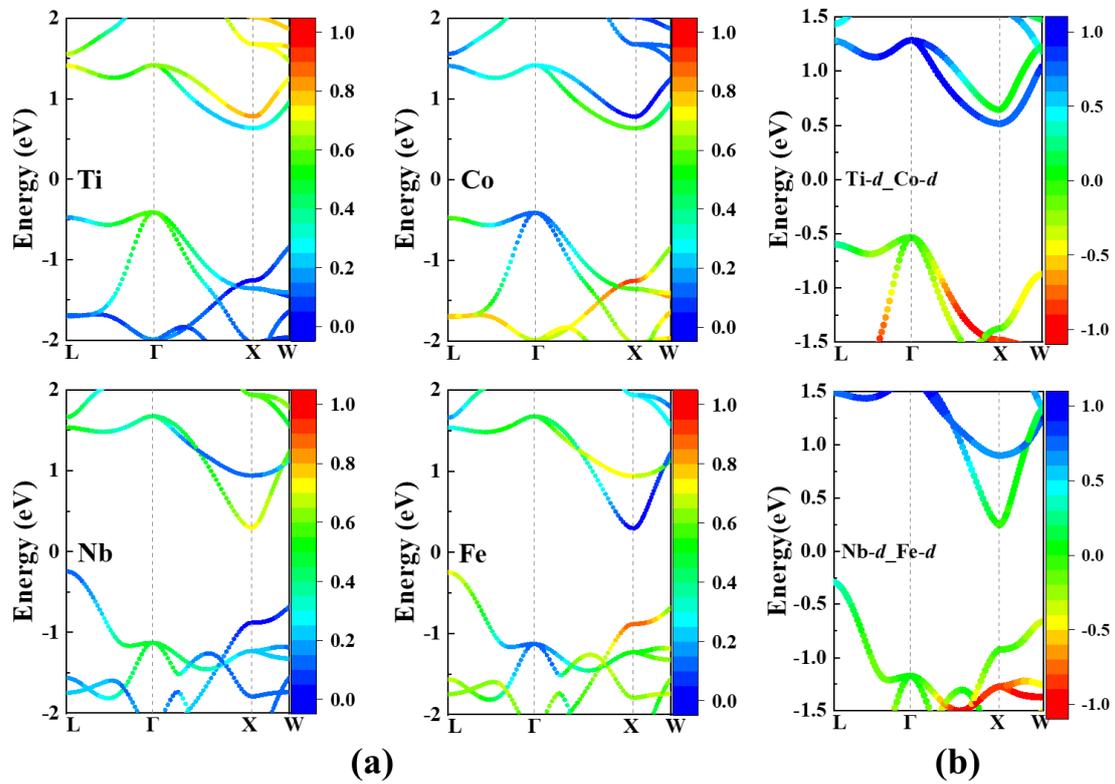

**Figure S4.** (a) Projected band structure of the primitive cell on Ti/Co for TiCoSb and Nb/Fe for NbFeSb, repectively. (b) Band-resolved projected crystal orbital Hamilton (pCOHP) of the primitive cell on Ti's *d*-orbital and Co's *d*-orbit for TiCoSb as well as Nb's *d*-orbital and Fe's *d*-orbit for NbFeSb, respectively, the positive values of pCOHP denote the antibonding states, and the negative values denote the bonding states.

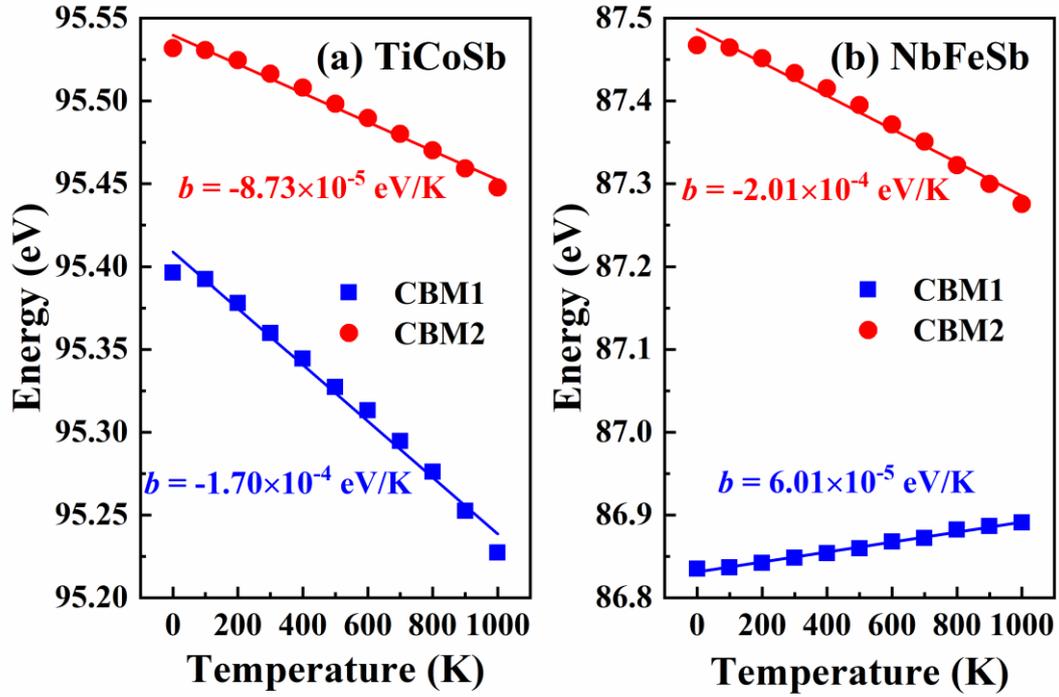

**Figure S5.** Absolute energy level shift for CBM1 and CBM2 with temperature in (a) TiCoSb and (b) NbFeSb, respectively. The average value of the first valence electron energy band as the reference level. The linear fitting parameter *b* is shown.

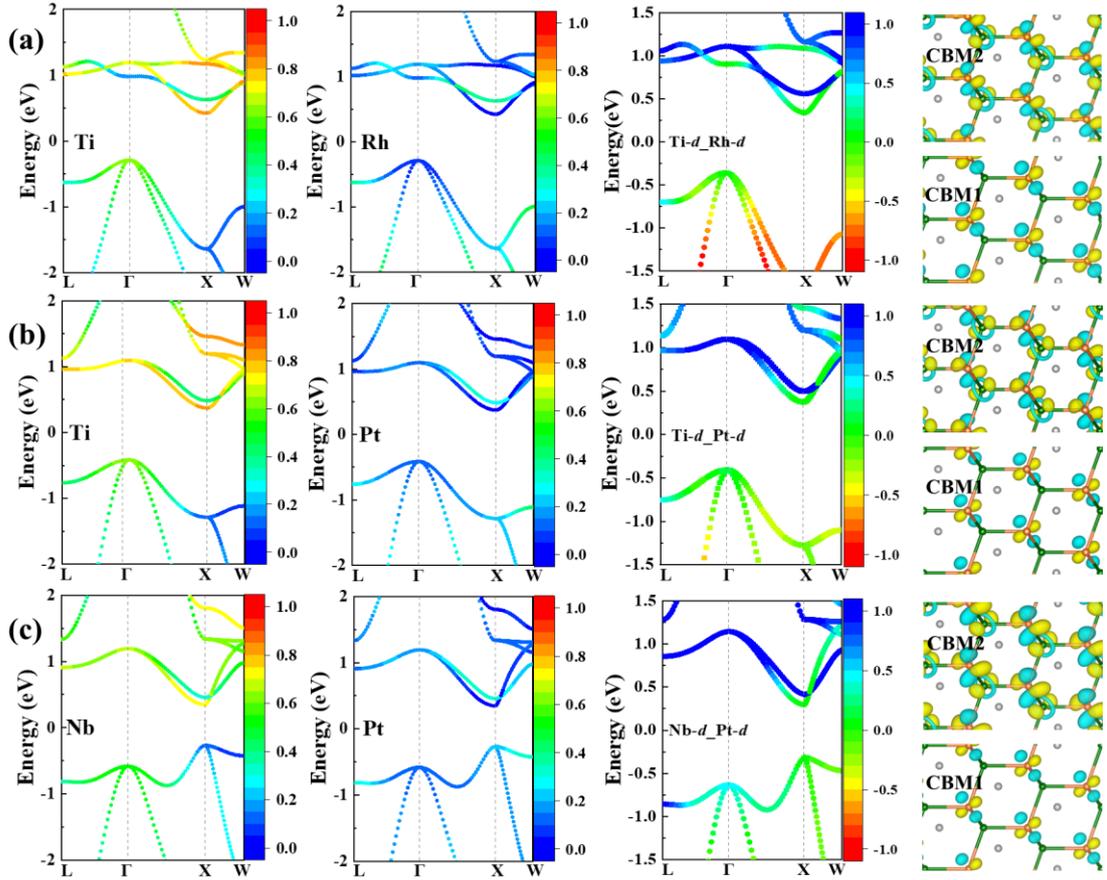

**Figure S6.** Projected band structure, band-resolved projected crystal orbital Hamilton (pCOHP) and wavefunctions of CBM1 and CBM2 in primitive cell for (a) TiRhBi, (b) TiPtSn and (c) NbPtTl, respectively. The positive values of pCOHP denote the antibonding states, and the negative values denote the bonding states.

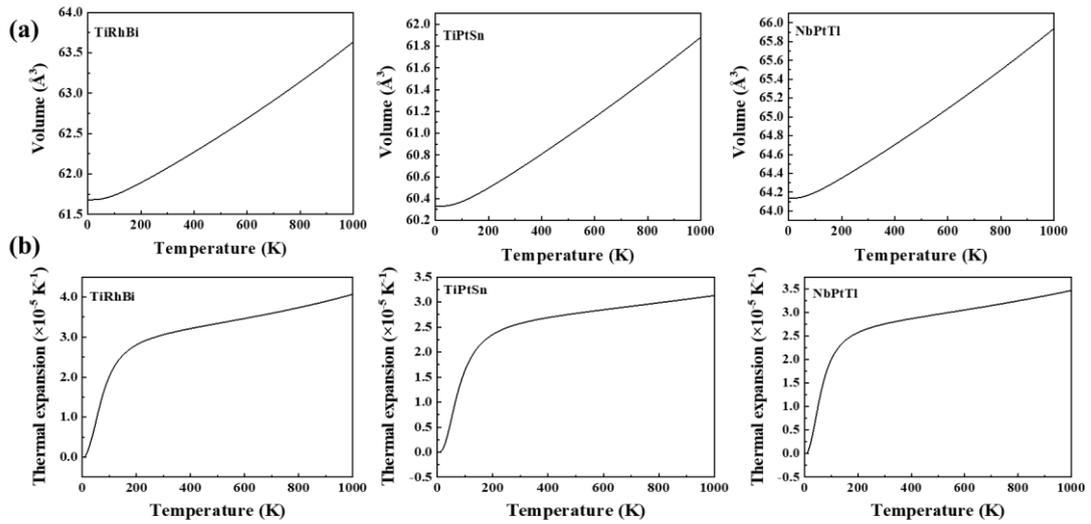

**Figure S7.** (a) Volume-temperature and (b) volumetric thermal expansion coefficient-temperature relations in TiRhBi, TiPtSn and NbPtTl, respectively.

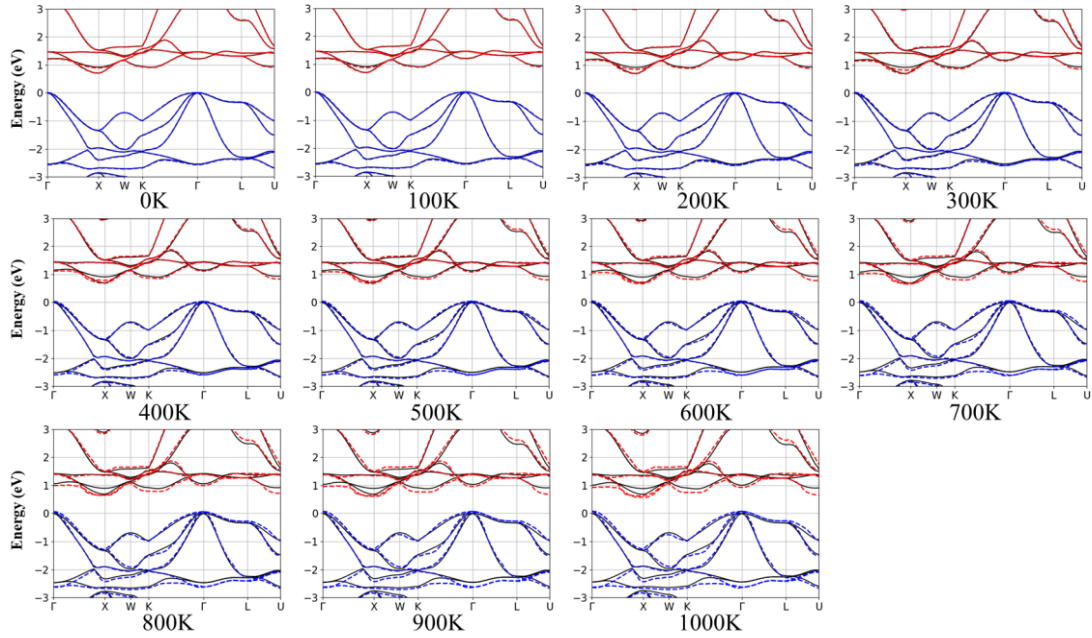

**Figure S8.** Temperature-dependent band structure of TiRhBi from 0 K to 1000 K. The lattice expansion and vibration effects are both considered. The black line is obtained without temperature effect, the blue and red dashed lines are valence and conduction bands at a given temperature, respectively.

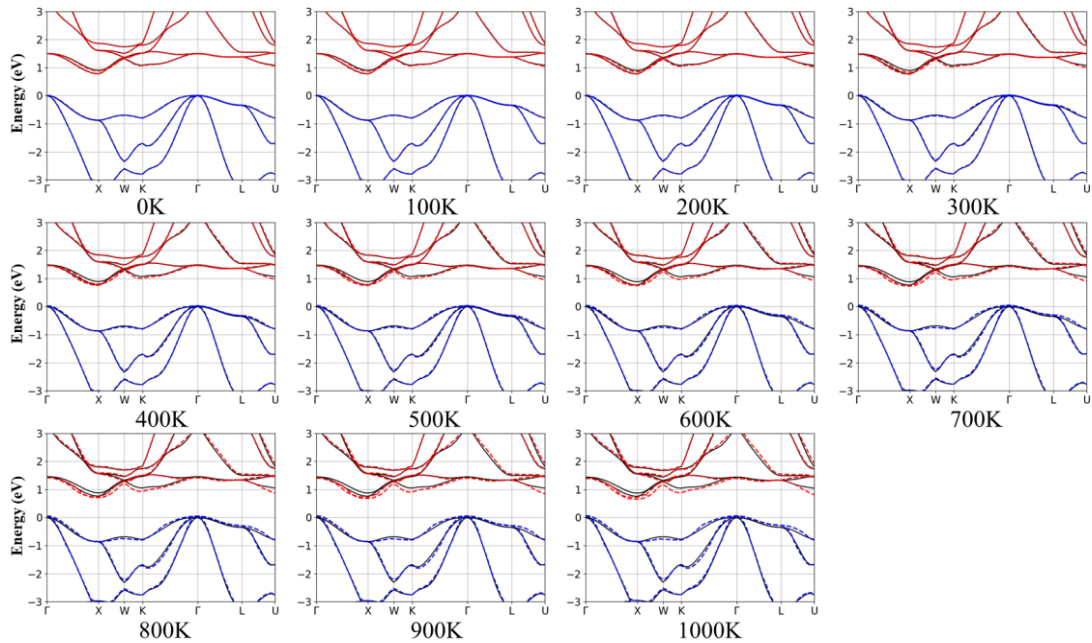

**Figure S9.** Temperature-dependent band structure of TiPtSn from 0 K to 1000 K. The lattice expansion and vibration effects are both considered. The black line is obtained without temperature effect, the blue and red dashed lines are valence and conduction bands at a given temperature, respectively.

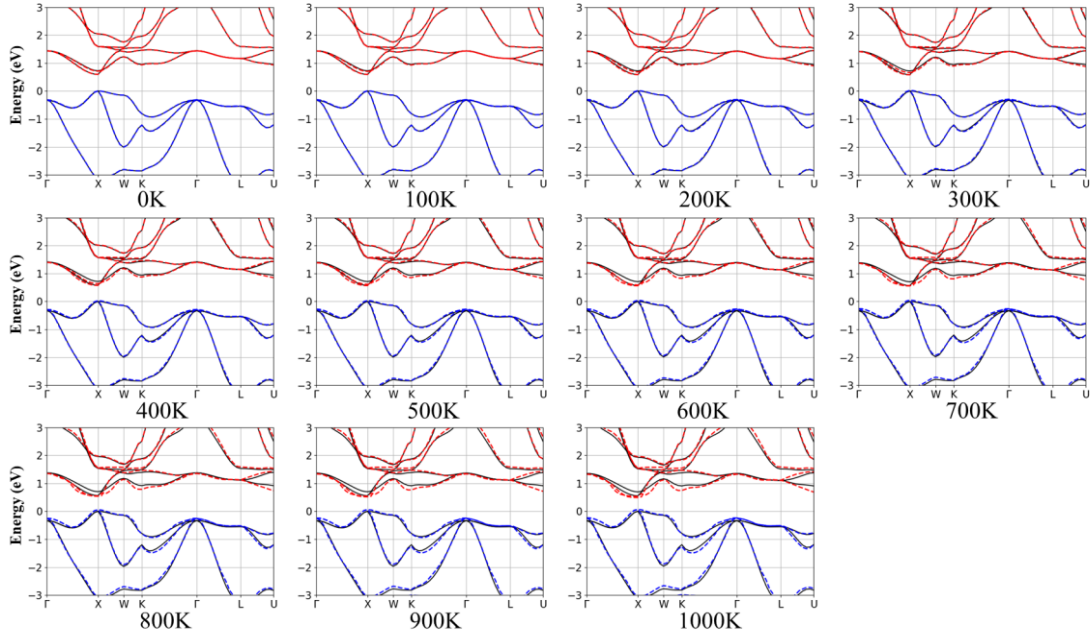

**Figure S10.** Temperature-dependent band structure of NbPtTl from 0 K to 1000 K. The lattice expansion and vibration effects are both considered. The black line is obtained without temperature effect, the blue and red dashed lines are valence and conduction bands at a given temperature, respectively.

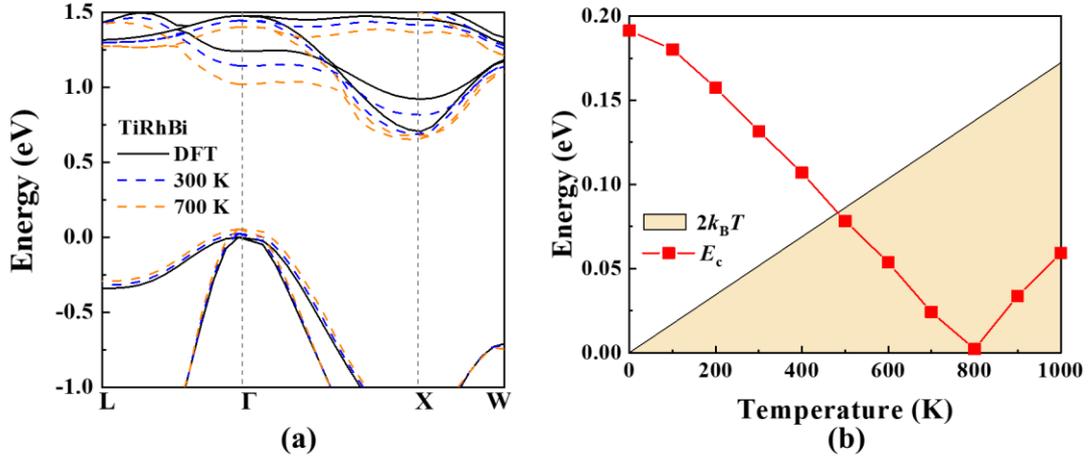

**Figure S11.** (a) Band structure of TiRhBi without temperature effect (DFT, black line), at 300 K (blue-dashed line), and at 700 K (orange-dashed line), respectively. (b) Temperature dependence of the energy difference $E_c$ between CBM1 and CBM2 for TiRhBi. The black line represents a Fermi distribution with $2k_BT$.

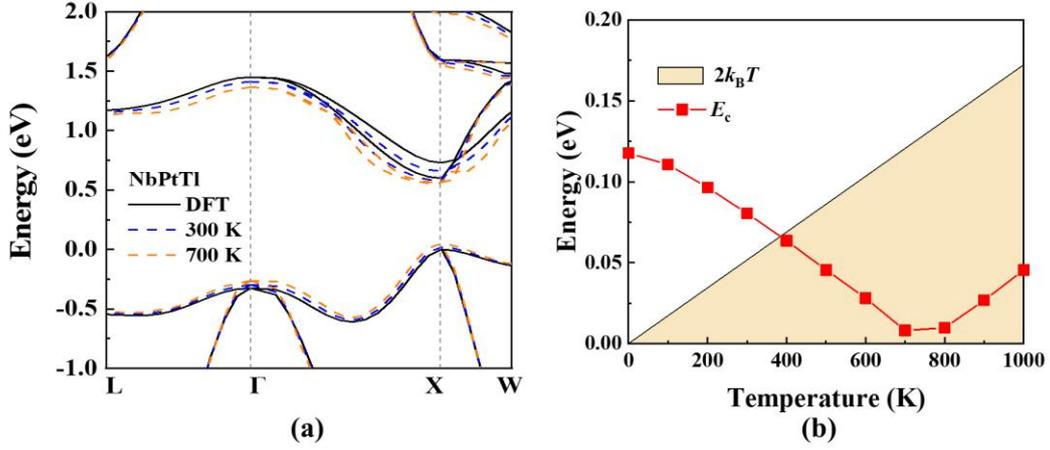

**Figure S12.** (a) Band structure of NbPtTl without temperature effect (DFT, black line), at 300 K (blue-dashed line), and at 700 K (orange-dashed line), respectively. (b) Temperature dependence of the energy difference $E_c$ between CBM1 and CBM2 for NbPtTl. The black line represents a Fermi distribution with $2k_BT$.

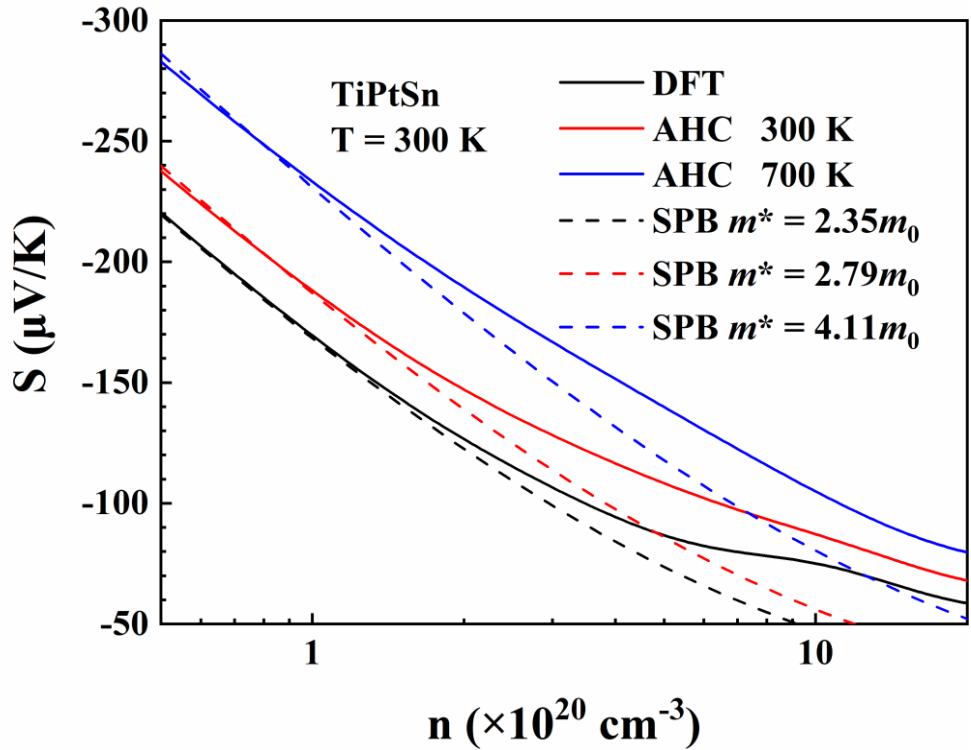

**Figure S13.** Calculated Seebeck coefficient $S$ at 300 K as the function of electron concentration in TiPtSn. Black, red and blue lines represent the $S$ obtained by using the band without temperature effect, at 300 K, and at 700 K, respectively. The corresponding effective mass obtained by the single parabolic band (SPB) model is also shown in dashed line.

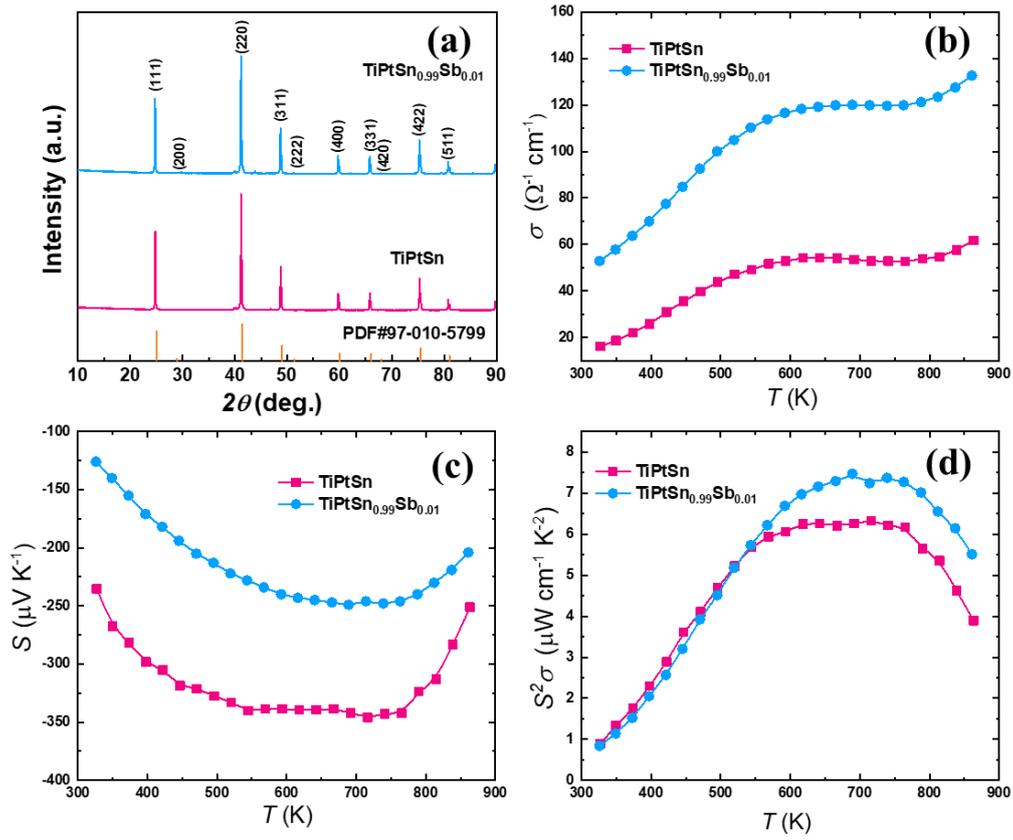

**Figure S14.** (a) XRD of TiPtSn and its 1% Sb doping TiPtSn$_{0.99}$Sb$_{0.01}$ samples. (b) Measured electrical conductivity $\sigma$, (c) measured Seebeck coefficient $S$ and (d) power factor $S^2\sigma$ as the function of temperature $T$ in TiPtSn and 1% Sb doped TiPtSn$_{0.99}$Sb$_{0.01}$, respectively.

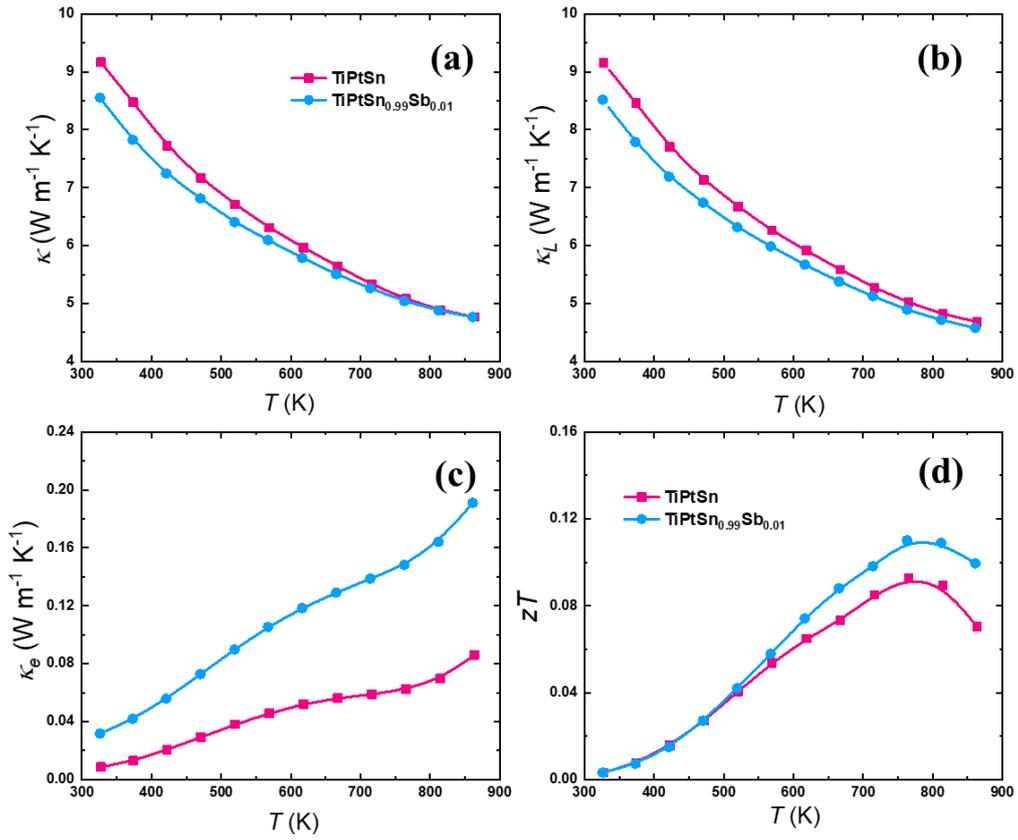

**Figure S15.** (a) Total thermal conductivity $\kappa$, (b) measured lattice thermal conductivity $\kappa_L$, (c) measured electrical thermal conductivity $\kappa_e$ and (d) figure of merit $ZT$ as the function of temperature $T$ in TiPtSn and 1% Sb doped TiPtSn$_{0.99}$Sb$_{0.01}$, respectively.

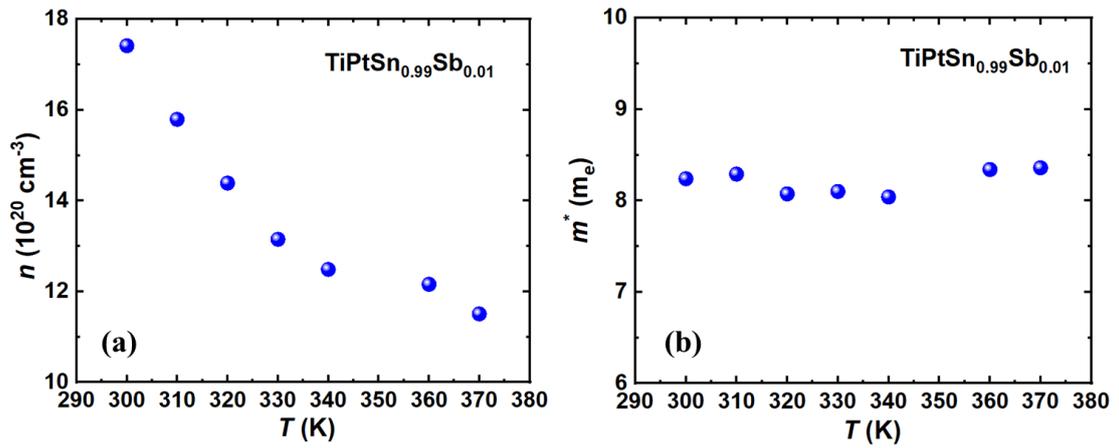

**Figure S16.** (a) Experimental electron concentration and (b) corresponding Hall effective mass as the function of temperature $T$ in 1% Sb doped TiPtSn$_{0.99}$Sb$_{0.01}$, respectively.

**Table S1.** Calculated deformation potential constant $D_{def}$ and Young's modulus $C$ at each temperature for TiPtSn.

| Temperature (K) | $D_{def}$ for VBM (eV) | $D_{def}$ for CBM (eV) | $C$ (GPa) |
|---|---|---|---|
| 0 | 2.550 | 1.553 | 136.112 |
| 100 | 2.554 | 1.555 | 136.421 |
| 200 | 2.549 | 1.54 | 137.433 |
| 300 | 2.541 | 1.532 | 136.248 |
| 400 | 2.534 | 1.525 | 136.443 |
| 500 | 2.524 | 1.515 | 135.903 |
| 600 | 2.518 | 1.508 | 135.936 |
| 700 | 2.515 | 1.503 | 137.467 |
| 800 | 2.513 | 1.498 | 138.850 |
| 900 | 2.492 | 1.481 | 136.595 |
| 1000 | 2.483 | 1.472 | 134.204 |